\documentclass[%
 reprint,
superscriptaddress,
 amsmath,amssymb,
 aps,
]{revtex4-2}

\usepackage{graphicx}
\usepackage{dcolumn}
\usepackage{bm}
\usepackage{svg}
\usepackage{amsmath}
\usepackage{mathtools}
\usepackage{dsfont}
\usepackage{xr}
\begin{document}

\preprint{APS/123-QED}

\title{A New Framework for Interfacial Statistics: Exact n-Point Correlations of Gaussian Level Sets}




\author{Aleksei M. Cherkasov}
\affiliation{Center for Computational Physics, Landau School of Physics and Research, Moscow Institute of Physics and Technology, 9 Institutskiy per., Dolgoprudny, 141701, Moscow Region, Russia}

\author{Kirill M. Gerke}
\affiliation{Center for Computational Physics, Landau School of Physics and Research, Moscow Institute of Physics and Technology, 9 Institutskiy per., Dolgoprudny, 141701, Moscow Region, Russia}
\affiliation{Schmidt Institute of Physics of the Earth of Russian Academy of Sciences, Bolshaya Gruzinskaya 10, Moscow, 123242, Russia}

\author{Aleksey Khlyupin}
\affiliation{Center for Computational Physics, Landau School of Physics and Research, Moscow Institute of Physics and Technology, 9 Institutskiy per., Dolgoprudny, 141701, Moscow Region, Russia}
\affiliation{Phystech School of Applied Mathematics and Computer Science, Moscow Institute of Physics and Technology, Institutskiy lane 9, Dolgoprudny, Moscow region, 141700, Russia}
\date{\today}

\begin{abstract}
We derive exact analytical expressions for higher-order correlations of Gaussian level-set interfaces, establishing a direct link between bulk field statistics and interface geometry. This framework enables efficient reconstruction of disordered media, detailed modeling of {Gaussian foams}---structured, double-thresholded interfaces with tunable morphology---and analysis of {memory-enhanced anisotropy} reflecting directional persistence in surface structure. These results open new possibilities for characterizing complex transport, guiding stochastic reconstructions, designing materials with desired properties, quantifying the information content of correlation functions and modelling directional processes on irregular boundaries.

\end{abstract}


\thanks{Published in Phys. Rev. Lett. 136, 196101 (2026)}

\maketitle{}


\section{Introduction}

From quantum field theory to cosmology, condensed matter physics, geostatistics, and materials science, Gaussian Random Fields (GRFs) and correlation functions are fundamental tools for describing spatial fluctuations in disordered systems \cite{Torquato_book,roberts1997statistical,tahmasebi2013cross,lavrukhin2023measuring,zubov2024search}. A key challenge is the analytical characterization of interface correlations, which describe surface physics, encompassing fluid--surface interactions \cite{khlyupin2017random,guskov2025nanodroplets,nesterova2025role}, scattering from surface-supported nanoparticles \cite{gommes2019small}, and the overall morphology crucial to physical properties \cite{Torquato_book,Sahimi_book}. While numerical methods can approximate these functions \cite{malmir2018higher,postnicov2024evaluation}, they do not typically yield closed-form expressions valid across different stochastic media, making it hard to derive exact links between bulk properties and interfaces.

Here, we develop an exact analytical framework for $n$-point interface correlation functions in truncated GRFs --- a class of models where a continuous field is thresholded to define complex, irregular interfaces \cite{hristopulos2020random}. We establish a direct link between the two-point correlation of the original field, its truncated version, and surface statistics, overcoming a long-standing limitation in stochastic geometry. Our results show that the surface correlation structure can be explicitly expressed in terms of the bulk field, enabling precise analytical predictions for interfacial properties. This provides a new theoretical foundation for stochastic media reconstruction \cite{adler1990flow,roberts1997statistical,yeong1998reconstructing,tahmasebi2013cross,karsanina2018hierarchical}, making it possible to model physically meaningful structural features and approach information-based assessment of descriptor quality \cite{cherkasov2024towards}.

Moreover, obtained interface correlation functions can serve as genuine correlation functions for Gaussian foams --- ribbon-like structures formed by thresholding at $\alpha \pm \epsilon$ unlike plurigaussian modelling \cite{roberts1995transport}. These morphologies can exhibit intriguing properties, including bicontinuity, enhanced mechanical strength, and energy absorption capacity \cite{yang2022high}. This enables their analytical characterization and provides a powerful tool to guide the design of lightweight and mechanically robust architectures in materials science. Furthermore, higher-order interface correlations reveal directional memory effects, allowing boundaries to be described as random walks \cite{gorelik2014quantitative, deng2011atomistic, bochicchio2017into} with non-trivial memory, a crucial insight for modeling surface transport and reactions.
\begin{figure}
\centering
\includegraphics[width=1\linewidth]{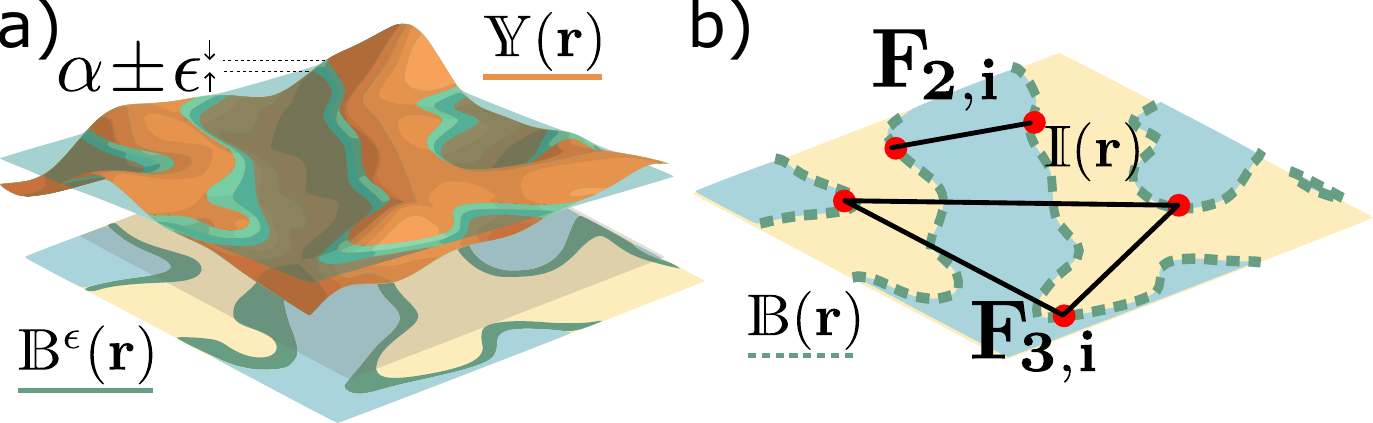}
\caption{
Schematic representation of the Gaussian field $Y(\mathbf{r})$, which, after thresholding at $\alpha$, gives binary field $\mathbb{I}(\mathbf{r})$ with finite-thickness $\epsilon$-regularized boundary $B^{\epsilon}(\mathbf{r})$. The limit field is $B(\mathbf{r}) = \lim_{\epsilon \to 0} B^{\epsilon}(\mathbf{r})/2\epsilon$. Correlations $F_{2,i}$ and $F_{3,i}$ depend on relative displacements.
}

\label{fig:evalf}
\end{figure}

\section{General model for n-point interface correlation function}
We start with the generation and characterization of a correlated, uniform, isotropic Gaussian random field $Y(\mathbf{r})$ with unit standard deviation, which is achieved through the convolution of white noise $\xi(\mathbf{r})$ with a symmetric kernel $h(\mathbf{r})$ in $N$-dimensional space:
\begin{equation}
\mathds{Y}(r) = \int_{R^N}^{}\xi(r-t)h(t)d^Nt
\label{eq:y}
\end{equation}

Autocorrelation function
$
    G_2(r_1, r_2) = \langle \mathds{Y}(r_1) \mathds{Y}(r_2) \rangle
$
due to the stationarity of the field $\mathds{Y}(r)$ depends only on the relative displacement $\tau = r_1 - r_2$ between field points. Substituting Eq.~\eqref{eq:y} into the definition of $G_2$ and utilizing the properties of white noise, we obtain:
\begin{align}
    G_2(\tau) =  \int_{R^N}^{}h(s)h(s - \tau)d^Ns
    \label{eq:g2def}
\end{align}

Thresholding $\mathds{Y}(r)$ at level $\alpha$ generates the binary field $\mathds{I}_\alpha(r) = 1$ if $\mathds{Y}(r) > \alpha$, and $\mathds{I}_\alpha(r) = 0$ otherwise.

In classical treatments \cite{Torquato_book, ma2018precise}, the interface is constructed from the gradient magnitude of a random-phase indicator function, yielding a boundary of fixed thickness. In contrast, within the level-set framework \cite{gommes2019small}, we define the interface as the region enclosed between the level sets $\alpha \pm \epsilon$, thereby introducing an $\epsilon$-regularized boundary of spatially varying thickness,
\begin{equation}
      \mathds{B}^\epsilon(r) = \mathds{I}_{\alpha - \epsilon}(r) - \mathds{I}_{\alpha + \epsilon}(r).
      \label{eq:b2}
 \end{equation}
 
Accordingly, instead of restricting the analysis to the constant-thickness surface correlation function \cite{Torquato_book}, we introduce a more general class of interface correlation functions based on the $\epsilon$-regularized boundary field.

Specifically, we define
\begin{equation}
F_{n,i}^\epsilon(r_1, \dots, r_n)
= \frac{1}{(2\epsilon)^n}\left\langle\,\mathds{B}^\epsilon(r_1)\dots \mathds{B}^\epsilon(r_n) \right\rangle,
\end{equation}
where $\langle \cdot \rangle$ denotes ensemble averaging. 

In the limiting case $\epsilon \to 0$, the regularized boundary reduces to a sharp, zero-thickness interface with non-uniform local weighting, given by the Dirac delta function (see Supplemental Material~\cite{SM} Sec. I\thinspace A),
\begin{equation}
\lim_{\epsilon \to 0}\frac{1}{2\epsilon}\mathds{B}^\epsilon(r) = \mathds{B}(r) \equiv \delta(\mathds{Y}(r) - \alpha).
\end{equation}
Accordingly, the $n$-point interface correlation function trends to
\begin{align}
\lim_{\epsilon \to 0} F_{n,i}^\epsilon(r_1, \dots, r_n)=\nonumber\\
F_{n,i}(r_1, \dots, r_n)
\equiv \langle \mathds{B}(r_1)\dots \mathds{B}(r_n) \rangle.
\label{eq:fns}
\end{align}

Spectral representation of the Dirac delta function $\delta(\mathds{Y}(r)-\alpha) = \frac{1}{2\pi}\int_{-\infty}^{+\infty}e^{i(\mathds{Y}(r)-\alpha)x}dx$ transforms definition for $F_{n,i}(\tau)$ (Eq. \ref{eq:fns}) to the expression with explicitly given $\mathds{B}(r)$ (Eq. \ref{eq:b2}):
\begin{equation}
F_{n,i}(\tau) =\frac{1}{(2\pi)^n}\prod_i\int_{-\infty}^{+\infty}dx_i e^{-i\alpha \sum_{i=1}^n x_i} \langle e^{i\sum_{i = 1}^n \mathds{Y}(r_i) x_i}\rangle.
\end{equation}
Then, we introduce a variable change: $e^{i\kappa}=e^{i\sum\mathds{Y}(r_i) x_i}
$
with $\mathds{Y}(r)$ evaluated in Eq. \ref{eq:y}:
\begin{align}
\kappa = \sum_{i = 1}^n x_i \int_{-\infty}^{+\infty}\xi(r_i - t)h(t)dt =\int_{-\infty}^{+\infty}\xi(t)\beta(t)dt
\end{align}
with $\beta(t) = \sum_{i=1}^n x_i h(r_i - t)$.
Using the properties of white noise $\xi(t)$, we obtain:
\begin{eqnarray}
\langle e^{i\kappa} \rangle =e^{-\frac{1}{2}\int_{-\infty}^{\infty}\beta^2(t)dt}
\end{eqnarray}
This product can be expressed as:
\begin{equation}
\int_{-\infty}^{+\infty}\!\!\!\!\!\beta^2(t)dt = \sum_{i,j}x_ix_j\int_{-\infty}^{+\infty}\!\!h(r_i-t)h(r_j-t)dt
=\mathbf{x} \cdot \!A_n \cdot \mathbf{x^T}
\end{equation}
where $\mathbf{x} = (x_1, \dots, x_n)^T$ and matrix $A_n$ elements are autocorrelation function values for different pairs of points $[A_n]_{ij} = G_2(r_i, r_j) = G_2(r_i - r_j)$ (see Eq. \ref{eq:g2def}). We use variable change $J = -\alpha
\begin{pmatrix}
1 &\dots &1
\end{pmatrix}^T
$ in common multivariate Gaussian integral to obtain the final expression:
\begin{align}
\label{eq:fnsfin}
F_{n,i}(\tau) &= \frac{1}{(2\pi)^n} \prod_i \int_{-\infty}^{+\infty} \!\!dx_i 
\exp\!\left(-i\alpha J\mathbf{x} - \frac{1}{2} \mathbf{x} \cdot A \cdot \mathbf{x^T} \right) \nonumber\\
&= \frac{1}{(2\pi)^{n/2} \sqrt{|A_n|}} 
\exp\!\left(-\frac{\alpha^2}{2} \sum_{i,j} [A_n^{-1}]_{ij} \right)
\end{align}

\section{2-point statistics: fast reconstruction by novel link between $G_2$, $S_2$ and $F_{2,i}$}

We now utilize the general model from Eq. \ref{eq:fnsfin} for the n-point interface correlation function to establish a relationship between the correlation of the truncated Gaussian field, $S_2$, its boundary, $F_{2,i}$, and the untruncated field correlation function, $G_2$. For the case of the two-point correlation function $A_n$ reduces to $2\times 2$ matrix: $[A_2]_{11} = [A_2]_{22} = 1$. $[A_2]_{12} = [A_2]_{12} = G_2(\tau)$.
The inverse matrix $A_2^{-1}$ takes the form:
\begin{equation}
A_2^{-1} = \frac{1}{
1 -G_2^2(\tau)
}
\begin{pmatrix}
1 & -G_2(\tau)\\
-G_2(\tau) & 1
\end{pmatrix}.
\end{equation}
The two-point interface correlation function, $F_{2,i}$, is fully defined by the Gaussian field correlation function, $G_2$:
\begin{equation}
     F_{2,i}(\tau) = \frac{1}{2\pi\sqrt{1 - G_2^2(\tau)}}\exp\left[-\frac{\alpha^2}{1 + G_2(\tau)}\right].
     \label{eq:f_ss}
\end{equation}
This expression is compatible with results from \cite{gommes2019small}, as shown in Supplemental Material~\cite{SM} Sec. I\thinspace A. Interestingly, it opens the way for information-based analysis of interface correlations, as in~\cite{cherkasov2024towards}.

\begin{figure*}
\centering
\includegraphics[width=\textwidth]{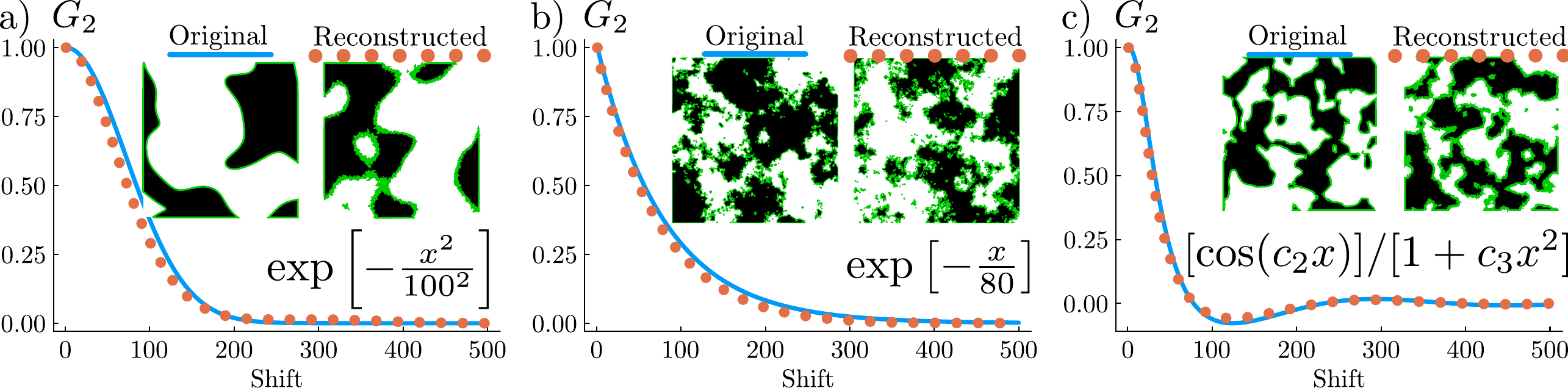}
\caption{
The figures show three correlation functions: Gaussian (a), exponential (b), and Gaussian oscillating (c). Both analytical and numerical evaluations demonstrate strong agreement, with minor discrepancies attributed to numerical evaluation. The insets display examples of binary fields. The left column shows fields generated from original analytical $G_2$ correlation functions, while the right column depicts reconstructions using numerically evaluated $G_2$. The reconstructed fields exhibit noisier boundaries due to numerical $G_2$ assessment.}
\label{fig:compar_cf}
\end{figure*}

From the other side, the autocorrelation function $S_2(\tau) = \langle \mathds{I}(r) \mathds{I}(r+\tau)\rangle$ of the binary field $\mathds{I}(r)$ is given in \cite{roberts1997statistical, berk1987scattering, berk1991scattering, teubner1991level} as:
\begin{equation}
S_2(\tau) = \frac{1}{2\pi}\int_0^{G_2(\tau)}\!\!\!\!\exp\left[-\frac{\alpha^2}{1 + t}\right]\frac{dt}{\sqrt{1 - t^2}} + p^2
\label{eq:s2}
\end{equation}
where the ratio of binary positive fraction $\mathds{I}(r)$ is $p=(2 \pi)^{-1 / 2} \int_{-\infty}^\alpha e^{-t^2 / 2} dt$. We observe that:
\begin{equation}
\frac{\partial }{\partial G_2}S_2\left[G_2(\tau)\right] = F_{2,i}(\tau)
\end{equation}
is a part of composed function derivative:
\begin{equation}
\frac{\partial G_2}{\partial \tau} = \frac{\partial S_2}{\partial \tau}\cdot\left({\frac{\partial S_2}{\partial G_2}}\right)^{-1} = \frac{1}{F_{2,i}}\frac{\partial S_2}{\partial \tau}
\label{eq:comp_der}
\end{equation}

Integrating Eq.~\ref{eq:comp_der} assuming $G_2(0) = 1$ naturally yields an explicit expression for $G_2$, formulated in terms of the autocorrelation function $S_2$ of the binary field and the two-point interface correlation function $F_{2,i}$:
\begin{equation}
G_2(\tau) = 1 + \int_{0}^{\tau}\frac{1}{F_{2,i}(x)}\frac{d}{dx}S_2(x)dx
\label{eq:G_fin}
\end{equation}
It is intriguing that the correlation function of three interconnected yet distinct fields, differing in spatial scales and curve shapes, consistently demonstrates a functional dependency.
Retrieving physical properties of heterogeneous structures often involves stochastic reconstruction to generate digital twins that replicate key characteristics of the original material~\cite{tahmasebi2013cross, rozman2002uniqueness}. Reconstruction approaches commonly rely on the two-point correlation function $S_2$, which can be obtained not only from direct 2D or 3D imaging but also from small-angle X-ray scattering (SAXS) measurements~\cite{gommes2019small}. In the latter case, the presence of surface-supported nanoparticles leads to an additional contribution from the surface correlation function $F_{ss}$ in the scattering intensity, that can provide more information about interfacial correlation properties. 

Existing methods, such as optimization-based approaches~\cite{yeong1998reconstructing, karsanina2018hierarchical} and Gaussian field truncation~\cite{roberts1997statistical}, reproduce $S_2$ and may also incorporate $F_{ss}$~\cite{jiao2009superior}, but they rely on complex inverse procedures or specific approximations. 
In contrast, Eq.~\ref{eq:G_fin} links $G_2$ to bulk and interfacial correlation information. Although Eq.~\ref{eq:G_fin} is naturally formulated in terms of the $F_{2,i}$ interface correlation function introduced here, for truncated Gaussian media we find that, at the two-point level, it exhibits behavior closely similar to the conventional surface correlation function $F_{ss}$. In practice, we therefore use $F_{ss}$ as an input descriptor, while keeping the distinction between the two definitions explicit.

To demonstrate this, we perform a series of reconstructions on 2D structures generated from Gaussian fields truncated at $\alpha = -0.1$, using Gaussian ($\sim \exp[-x^2/100^2]$), exponential ($\sim \exp[-x/80]$), and oscillating $\sim \cos(0.0204 x)/(1+0.025 x^2)$ $G_2$ correlation functions. The original structures yield $S_2$ and $F_{ss}$, which are subsequently used to compute the reconstructed $G_2$ and generate replicas via Eq.~\ref{eq:G_fin}. A comparison between the original and reconstructed samples, along with their correlation functions (Fig.~\ref{fig:compar_cf}), demonstrates good agreement. Minor discrepancies arise from the numerical evaluation of $S_2$ and $F_{2,i}$, as well as from the assumptions introduced when substituting $F_{2,i}$ by $F_{ss}$. More details on the reconstruction procedure can be found in Supplemental Material~\cite{SM} Sec. I\thinspace B.
 \begin{figure*}
    \centering
    \includegraphics[width=0.95\linewidth]{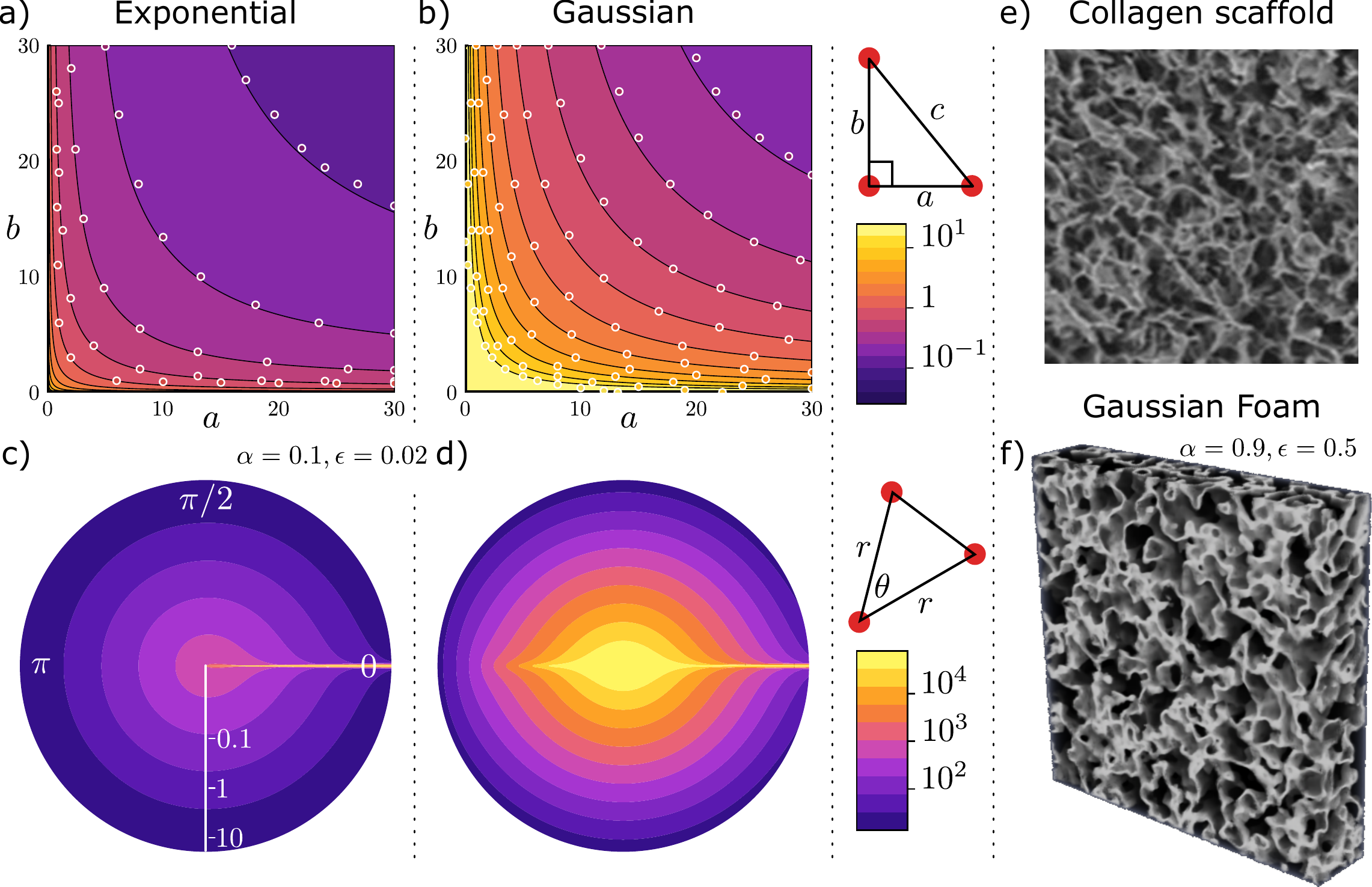}
    \caption{(a) Three-point interface correlation function $F_{3,i}(a,b,c=\sqrt{a^2+b^2})$ for an exponential $G_2$ field. (b) Same function for a Gaussian $G_2$ field. Scatter: numericdal $F_{3,i}^{\epsilon}$ using CorrelationFunctions.jl \cite{postnicov20232}. (c) Polar plot of $F_{3,i}(r,r,\theta)$ for the exponential $G_2$ field. (d) Corresponding plot for the Gaussian $G_2$ field. (e) SEM image of a real collagen scaffold \cite{szychlinska2020evaluation}. (f) Our Gaussian Foam reconstruction with Matern-2 correlation, illustrating visual consistency with collagen structures.}
    \label{fig:angle-hm-collagen}
\end{figure*}
\section{Higher-order Interface Correlation Functions and Applications}
Higher-order surfacial correlations play a crucial role in characterizing complex interfaces, especially when two-point functions fail to capture higher-order connectivity or directional anisotropy \cite{Torquato_book}. Among them, the three-point function serves as a key descriptor, providing additional insights into the spatial arrangement and angular correlations of surface elements.

In this particular case, the matrix $A_3$ in Eq.~\ref{eq:fnsfin} is a $3 \times 3$ matrix with diagonal entries $[A_3]_{ii} = 1$ for $i \in \{1, 2, 3\}$, and off-diagonal entries defined as $[A_3]_{12} = [A_3]_{21} = G_2(a)$, $[A_3]_{13} = [A_3]_{31} = G_2(b)$, and $[A_3]_{23} = [A_3]_{32} = G_2(c)$, where $a$, $b$, and $c$ denote the sides of the triangle formed by the three points. This explicit formulation enables an analytical evaluation of the three-point function $F_{3,i}$. Importantly, this expression satisfies the limit cases for two close and two remote points, as discussed in the Supplemental Material~\cite{SM} Sec. II\thinspace A, thereby demonstrating the internal consistency of the model and capturing the singularity behavior in asymptotic configuration.

We observe a good match between the analytically evaluated $F_{3,i}$, shown as a heatmap with isolines in Fig.~\ref{fig:angle-hm-collagen}a,b, and the numerically evaluated result $F_{3,i}^{\epsilon}$, presented as scatter points in the same figure. These results are obtained for $\epsilon$-regularized boundary $B^\epsilon$ in a right-angled triangle configuration (see Supplemental Material~\cite{SM} Sec. II\thinspace B for details). This agreement not only confirms the internal consistency of the theoretical expression but, more importantly, demonstrates that it provides a robust reference for validating numerical packages and computational algorithms designed to evaluate higher-order surface statistics \cite{malmir2018higher,correlationfunctions}.

\subsection{Physical Properties of Gaussian Foams}
We denote the family of objects generated by Gaussian fields as Gaussian foams, which, in contrast to once-thresholded fields \cite{roberts1995transport} defined by \(Y > \alpha\), represent twice-thresholded ribbon-like structures forming layers between \(\alpha \pm \epsilon\), thus highlighting finite-thickness interfaces \cite{roberts1997statistical,gommes2019small}. Examples such as echinoderm stereom \cite{yang2022high} or collagen scaffolds are specific cases of this versatile class, as illustrated in Fig.~\ref{fig:angle-hm-collagen}e (SEM of collagen from \cite{szychlinska2020evaluation}) and Fig.~\ref{fig:angle-hm-collagen}f (our reconstruction using a Matérn-2 field \cite{matern1960spatial, cressie1999classes} thresholded at \(\alpha = 0.9 \pm \epsilon = 0.5\)), showing visual consistency.

Such structures can exhibit intriguing properties, including bicontinuity, smooth and continuous surface morphologies, as well as enhanced mechanical strength and energy absorption capacity, as highlighted in \cite{yang2022high}. In turn, to accurately predict macroscopic mechanical and transport properties, higher-order (n-point) correlation functions are essential \cite{Torquato_book, jiao2012quantitative}. Fortunately, our \(n\)-point correlations naturally serve this role for such structures. While approximations using correlation functions of order \(n \le 4\) are constructed for certain specific types of media \cite{Torquato_book}, our systems likely require higher \(n\). Although evaluating them numerically is challenging, our approach allows explicit analytical expressions for any order \(n\)-point correlation function of such Gaussian foams.  This connection opens a direct pathway from microstructural characterization to enabling informed design strategies for advanced heterogeneous structures. 

The intricate shape of the boundaries in such fields naturally leads to considering the analysis of possible trajectories along these surfaces. Some insights into this behavior can already be revealed by analyzing $F_i^{3}(r, r, \theta)$ as a function of side length and apex angle of an isosceles triangle — a function that also explicitly appears in evaluating transport properties, as shown in \cite{torquato1985effective,malmir2018higher}.

For different $G_2$ functions — exponential and Gaussian — two distinct boundary behaviors emerge (see Fig.~\ref{fig:angle-hm-collagen}c,d): exponential $G_2$ fields exhibit nearly circular isolines, with noticeable deviations only at small angles. In contrast, Gaussian $G_2$ fields exhibit enhanced intensity not only at small angles ($\theta \approx 0$) but also near $\theta \approx \pi$ when the radius is smaller than the correlation length. This onion-like structure indicates the presence of directional persistence \cite{gorelik2014quantitative}, suggesting that the boundary locally preserves preferred directions. At larger radii, these anisotropic effects gradually diminish, and the function approaches approximate circular symmetry, indicating a loss of correlation between steps along the interface.

\begin{figure}
\centering
\includegraphics[width=0.9\linewidth]{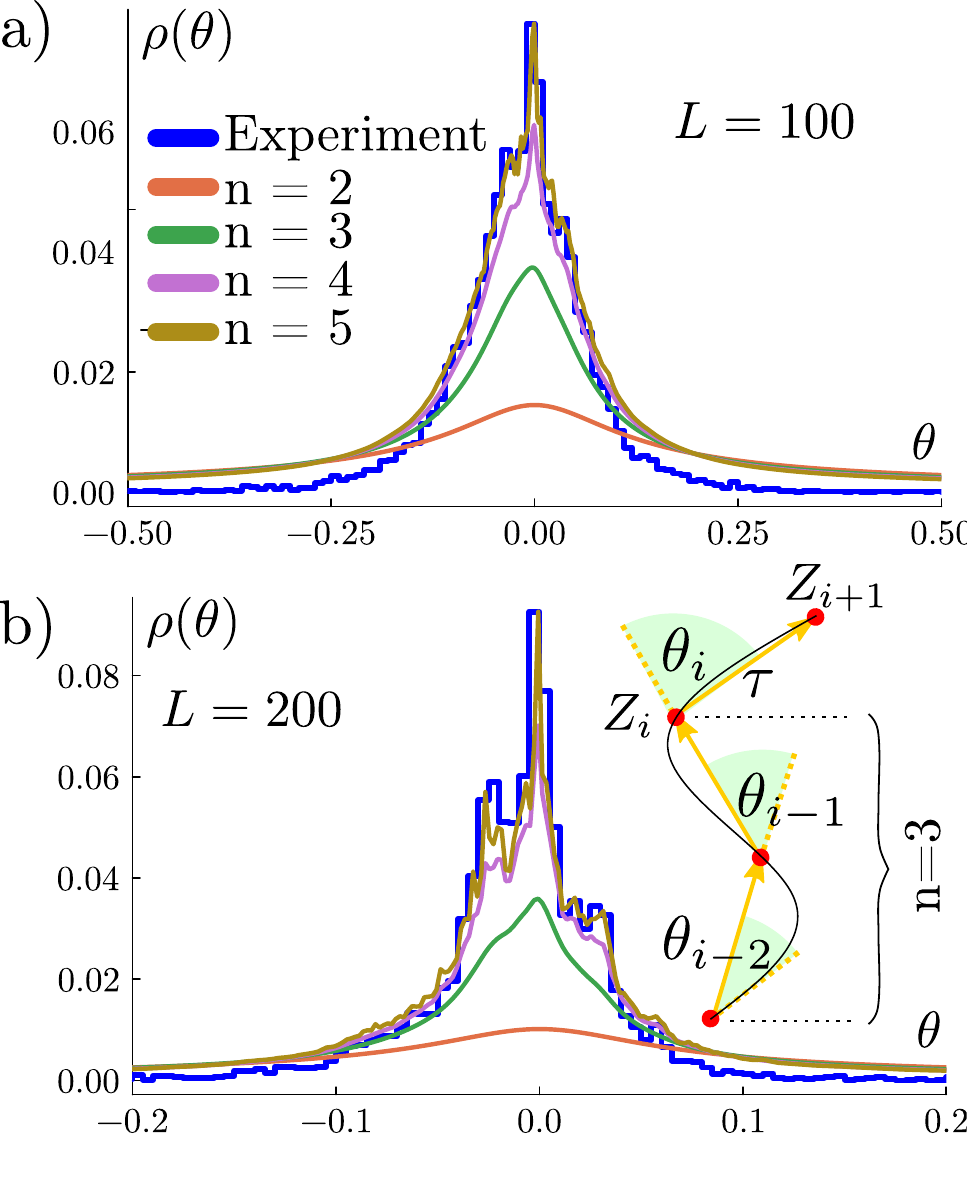}
\caption{Experimental and predicted conditional probability densities for two different correlation lengths, 100 (a) and 200 (b), using a step size of 5. The parameter $n$ indicates the number of previous points used for conditioning.}
\label{fig:angle_hists}
\end{figure}
\subsection{Memory-enhanced anisotropy}
\label{sec:rw}
Inspired by the directional persistence observed at the boundaries of Gaussian \( G_2 \) fields, we design a numerical experiment to study the trajectory of a hypothetical particle moving along a truncated field boundary with equal-length steps. Each step has a fixed size \( \tau \) and is characterized by a turning angle \( \theta_i \), capturing the local change in direction (see inset of Fig.~\ref{fig:angle_hists}).

Given the boundary’s smoothness, we hypothesize that turning angles can be fully described by the three-point interface correlation function \( F_{3,i}(\tau, \tau, \theta) \) (see Fig.~\ref{fig:angle-hm-collagen}d), similar to correlation between displacement vectors used in cell migration studies \cite{gorelik2014quantitative}.

Surprisingly, this assumption proves insufficient: knowledge of only two preceding points does not fully capture the observed behavior. This discrepancy reveals a striking phenomenon we refer to as memory-enhanced anisotropy—each turning angle \( \theta_i \) is influenced not only by the immediately preceding step but also by a longer sequence of prior steps, indicating a nontrivial directional memory embedded in the boundary dynamics.

To quantify this effect, we employ the conditional correlation function \cite{fisher1966statistical}, which gives the probability that a new point \( Z_{n+1} \) lies on the boundary, given \( n \) preceding points \( Z_1, \ldots, Z_n \):
\begin{equation}
P(Z_{n+1} \mid Z_1, \ldots, Z_n) \propto \frac{F_{n+1,i}(Z_1, \ldots, Z_{n+1})}{F_{n,i}(Z_1, \ldots, Z_n)}.
\label{eq:cnd_bound_dens}
\end{equation}
This approach directly builds on our main theoretical result (Eq.~\ref{eq:fnsfin}), capturing the intricate geometry of the boundary. 
We track the particle step-by-step, analytically compute the conditional distribution of each turning angle \( \theta_i \) from Eq.~\ref{eq:cnd_bound_dens}, and compare these predictions to empirical angle histograms obtained numerically from the recorded trajectory. As shown in Fig.~\ref{fig:angle_hists}, increasing the number of conditioning points $n$ leads to systematically better agreement between the predicted and observed distributions, confirming the significant role of trajectory memory in stochastic interphase transport.

Additional results for various correlation lengths and step sizes, along with a detailed account of the numerical methods, are presented in Supplemental Material~\cite{SM} Sec. III.
\section{Conclusion}

We present exact analytical expressions for $n$-point interface correlation functions in truncated Gaussian random fields, establishing an explicit link between bulk correlations, binary representations, and interfacial geometry. In contrast to classical approaches that formulate surface correlations for interfaces of constant thickness, the present framework derives interfacial statistics from the underlying field description, enabling explicit analytical relations between bulk structure and boundary morphology. This framework enables efficient reconstruction of disordered media, supports the modeling of Gaussian foams with tunable morphology, and captures memory-enhanced anisotropy in surface structures. Analytical and numerical results show strong agreement, confirming the robustness of the approach. These findings offer a unified theoretical basis for characterizing and reconstructing complex heterogeneous materials, with broad applications in stochastic transport, porous media analysis, and interfacial physics.
\section*{Acknowledgements}
The authors gratefully acknowledge support from the Ministry of Science and Higher Education of the Russian Federation (Agreement No. 075-03-2026-305 16.01.2026).

Authors sincerely thank anonymous reviewers, who provided extremely useful comments that helped in making this paper much better.

\nocite{*}
\bibliography{apssamp}

\end{document}